\documentclass[hyper]{JHEP3}
\usepackage{graphicx}

\input{epsf}
\usepackage{epsfig}
\usepackage{amssymb}
\usepackage{amsfonts}
\usepackage{amsbsy}
\usepackage[all]{xy}
\usepackage{amsmath}
\allowdisplaybreaks[2]

\usepackage{amssymb,amscd}
\usepackage{mathrsfs}
\usepackage{amsmath,amsthm}

\def\Tr{{\rm Tr}}

\def\IC{\mathbb{C}}

\def\IH{\mathbb{H}}

\def\IZ{{\mathbb{Z}}}

\def\CN {{\cal N}}

\def\half{\frac{1}{2}}

\def\one{{\hbox{ 1\kern-.8mm l}}}

\def\half{\frac{1}{2}}

\newcommand\be{\begin{equation}}
\newcommand\ee{\end{equation}}

\title{Mathieu twining characters for K3}

\author{Matthias R.~Gaberdiel, Stefan Hohenegger and
Roberto Volpato
\\ ~
\\
Institut f\"ur Theoretische Physik \\
ETH Zurich \\
CH-8093 Z\"urich \\
Switzerland }

\abstract{The analogue of the McKay-Thompson series for the proposed
Mathieu group action on the elliptic genus of K3 is analysed. The corresponding
NS-sector twining characters have good modular properties and satisfy remarkable
replication identities. These observations provide strong support for
the conjecture that the elliptic genus of K3 carries indeed an action of
the Mathieu group ${\mathbb M}_{24}$. }

\begin{document}

%%%%%%%%%%%%%%%%%%%%%%%%%%%%%%%%%%%%%%%%%%%%%%%%%%%%%%%%%%%%%%%%%%%%%

\section{Introduction and summary}

Recently, Eguchi, Ooguri and Tachikawa \cite{EOT} have observed that the
elliptic genus of K3 seems to carry an action of the Mathieu group ${\mathbb M}_{24}$.
More specifically, they noticed that the expansion coefficients of the elliptic genus of K3, when
written in terms of the elliptic genera of ${\cal N}=4$ superconformal representations,
can be expressed in terms of dimensions of Mathieu group representations. This intriguing
observation is very reminiscent of the famous observation of McKay and Thompson who found
that the Fourier expansion coefficients of the $J$-function can be written in terms of dimensions of
representations of the Monster group. This led to a development that is now usually
referred to as `Monstrous Moonshine', see \cite{Gannon}  for a nice review. One of the upshots
of this analysis is that the $J$-function can be thought of as the partition function of a
self-dual conformal field theory, the `Monster conformal field theory' \cite{Borcherds,FLM},
whose automorphism group is precisely the Monster group.

Unlike the situation of the Monster group, the dimensions of the Mathieu group are rather small,
and thus there is a certain amount of arbitrariness in the above observation regarding the elliptic
genus of K3. It would therefore be very interesting to subject this proposal to some more
stringent tests.  In the context of  Monstrous Moonshine, one of the tests concerned the
so-called McKay-Thompson series \cite{Thompson}. They are obtained from the
$J$-function by replacing the dimensions of the Monster group representations by their
characters with respect to a Monster group element. (In retrospect, they are the
`twining characters', where one inserts into the partition function of the Monster conformal
field theory a Monster group element.) It was observed early on by Conway \& Norton
that these McKay-Thompson series have nice modular properties and satisfy remarkable
replication identities \cite{CN}. The former identities have a natural interpretation in terms of
standard orbifold arguments \cite{Tuite:1990us}, while the latter can be obtained
by considering the twining characters for the symmetric power theories (as follows from
our analysis in section~\ref{sec:3.1}).
\smallskip

In this paper we study the analogues of the McKay-Thompson series for the case of K3,
{\it i.e.}\ the twining elliptic genera where we insert a Mathieu group element into the trace.
Using the proposed decomposition of the coefficients of the elliptic genus in terms of
Mathieu group representations, we can calculate the first few coefficients of these
twining genera explicitly. As for the McKay-Thompson series, standard orbifold arguments
suggest that they must have good modular properties. At least for the group elements of
small order, these modular constraints are very constraining, and we can use them to
determine the twining genera exactly. We furthermore show that they satisfy highly non-trivial
replication identities, relating twining genera corresponding to different group elements
to one another.
 The fact that all of this works out represents a very stringent consistency
check for the proposal of \cite{EOT}; indeed, our analysis implies that one of the
identifications of the coefficients in terms of Mathieu representation dimensions must
be different to what was proposed in \cite{EOT}. Our analysis thus gives strong
support to the idea that the elliptic genus of K3 carries indeed an action of the Mathieu group
${\mathbb M}_{24}$. It furthermore suggests that this is in some sense the natural
supersymmetric generalisation of Monstrous Moonshine to the case of ${\mathbb M}_{24}$.
(A different proposal for the analogue of Monstrous Moonshine for ${\mathbb M}_{24}$ was
made some time ago in the context of self-dual bosonic conformal field theories
by Dong \& Mason \cite{Dong:1994wm}.)

\medskip

The paper is organised as follows. In section~2 we recall some basic properties
of elliptic genera and their modular properties. We also review how the elliptic genus
of the symmetric power theory can be calculated in terms of the original elliptic genus.
In section~3 we briefly describe the proposal of \cite{EOT}, and then calculate the
twining elliptic genera and their associated NS twining characters. We explain in detail
which modular properties the latter should possess. Using this insight we then identify
the twining characters for group elements of small order in  terms of known theta
functions (or standard McKay-Thompson series), see (\ref{3.11}). In section~\ref{sec:3.1}
we explain what replication formulae one should expect, and demonstrate in a large
number of examples, that these identities are indeed satisfied for the twining characters
of K3. Our conclusions and avenues for future research are described in section~4.
We have relegated some of the more technical material to an appendix.

\section{Elliptic genera and their modular properties}

Let us begin by reviewing some standard material about elliptic genera and their Jacobi and modular
properties. The elliptic genus of an ${\cal N}=2$ superconformal algebra is defined by
\begin{equation}
\phi(\tau,z) = \hbox{Tr}_{{\cal H}_{RR}} \Bigl( q^{L_0-\frac{c}{24}} e^{2\pi i z J_0} (-1)^F \,
\bar{q}^{\bar{L}_0-\frac{\bar{c}}{24}} (-1)^{\bar{F}} \Bigr) \ .
\end{equation}
For the right-movers (whose modes are denoted by a bar), only the ground states contribute,
and hence the above expression is in fact independent of $\bar{q}$.
As is well-known \cite{Kawai:1993jk}, the  modularity properties of
conformal field theory together with spectral flow invariance and unitarity imply
that the elliptic genus is a \emph{weak Jacobi form} of index $m=\frac{c}{6}$
and weight zero \cite{EichlerZagier}.
A weak Jacobi form $\phi(\tau,z)$ of weight $w$ and index $m\in{\mathbb Z}$, with
$(\tau,z)\in \IH\times \IC$, satisfies the transformation laws
\begin{equation}\label{eq:jactmn1}
 \phi\Bigl(\frac{a \tau + b}{c \tau + d} , \frac{z}{c \tau + d}\Bigr) =
(c \tau+d)^w \, e^{ 2 \pi i m \frac{c z^2}{c \tau + d} } \, \phi(\tau,z)
\qquad \begin{pmatrix} a & b \\ c & d \\ \end{pmatrix} \in SL(2,\IZ) \ ,
\end{equation}
\begin{equation}\label{eq:jactmn2}
 \phi(\tau,z+ \ell \tau + \ell') = e^{-2 \pi i m
(\ell^2 \tau+ 2 \ell z)} \phi(\tau,z) \qquad\qquad\qquad
\ell,\ell'\in \IZ \ ,
\end{equation}
and has a Fourier expansion
\begin{equation}
 \phi(\tau,z) = \sum_{n \geq 0, \ell\in \IZ} c(n,\ell) q^n
y^\ell \end{equation}
with $c(n,\ell) = (-1)^w c(n,-\ell)$.
For the case of K3 that will concern us primarily in this paper, the elliptic genus equals
\cite{EOTY}
\begin{equation}\label{K3R}
\phi_{K3}(\tau,z)   =
2 y + 20 + 2 y^{-1} + q \Bigl(20 y^2 -128 y + 216 - 128 y^{-1} + 20 y^{-2} \Bigr) + {\cal O}(q^2) \ ,
\end{equation}
where $y= e^{2\pi i z}$. For the following it will be convenient to rescale the elliptic genus
by a factor of $\tfrac{1}{2}$, {\it i.e.}\ to define
\begin{equation}\label{K3Rr}
\phi_{\rm 1A}(\tau,z) = \frac{1}{2} \phi_{K3}(\tau,z) =
y + 10 +  y^{-1} + q \Bigl(10 y^2 - 64  y + 108 - 64 y^{-1} + 10 y^{-2} \Bigr) + {\cal O}(q^2) \ ,
\end{equation}
which is then directly equal to the standard weak Jacobi form
$\phi_{0,1}$.

To understand the analogy with the bosonic Monster conformal field theory, it is useful
to consider instead the NS-sector version of the elliptic genus, which is obtained by applying
one half unit of spectral flow to the left-movers, {\it i.e.} $l=l'=\frac{1}{2}$ in (\ref{eq:jactmn2});
the shift by $l=\frac{1}{2}$ is familiar from the spectral flow automorphism
\begin{equation}
L_0 \mapsto L_0 + l \, J_0 + l^2 \, \frac{c}{6} \ , \qquad
J_0 \mapsto J_0 +  l\, \frac{c}{3} \ ,
\end{equation}
while the shift by $l'=\frac{1}{2}$ is required so as to remove the $(-1)^F$ from
the trace, in order to get the standard NS-sector partition function without an insertion of $(-1)^F$.
The resulting NS-sector partition function will always be denoted by
\begin{equation}\label{NS}
\chi(\tau,z) = \exp\left[2 \pi i m \left(\frac{\tau}{4} + z+ \half \right) \right]
\phi\left(\tau, z + \frac{\tau}{2} + \half  \right) \ .
\end{equation}
Using the transformation properties of a Jacobi form it follows
easily that
\begin{equation}\label{eq:NSegtmrn}
\begin{split} \chi(-1/\tau, z/\tau) & = (-1)^m
e^{2\pi i\frac{m z^2}{\tau}}\, \chi(\tau,z)\\
\chi(\tau+2,z) & = (-1)^m\,  \chi(\tau,z) \ .\\
\end{split}
\end{equation}
In order to obtain a standard modular form (rather than a Jacobi form) we may
then put $z=0$; the resulting function (which
we shall simply denote by $\chi(\tau)$)  is non-trivial,
and it has simple transformation laws under the congruence
subgroup $\Gamma_\theta = \langle T^2, S\rangle$. For $m$ even we
have a strict modular function and for $m$ odd we have a function
with multiplier system given by $-1$ on the two generators \cite{Gaberdiel:2008xb},
\begin{equation}
\chi(-1/\tau) = (-1)^m \chi(\tau)\ , \qquad \chi(\tau+2) = (-1)^m \chi(\tau) \ .
\end{equation}
For the case of K3 it is explicitly given as \cite{Gaberdiel:2008xb}
\begin{equation}\label{K3NS}
\chi_{\rm 1A}(\tau) = \frac{1}{2} \chi_{K3}(\tau)
=  \kappa(\tau) =
\left(\frac{2 \vartheta_4(\tau)}{\vartheta_2(\tau)}\right)^2
-\left(\frac{2 \vartheta_2(\tau)}{\vartheta_4(\tau)}\right)^2
= q^{-\frac{1}{4}}(1- 20 \, q^{\frac{1}{2}} + \cdots ) \ .
\end{equation}
Our conventions regarding the theta functions are described in the appendix.

It should be noted that specialising to $z=0$ in the NS-sector does not
loose any information. Indeed, the NS character $\chi_{\rm 1A}(\tau)$
determines $\phi_{\rm 1A}(\tau,\tfrac{1}{2})$ by setting $z=0$ in
\begin{equation}\label{rel1}
 \chi_{\rm 1A}(\tau,z) = \phi_{\rm 1A}(\tau,0)\, \frac{\vartheta_4(\tau,z)^2}{\vartheta_2(\tau,0)^2}
-\phi_{\rm 1A}(\tau,\tfrac{1}{2})\frac{\vartheta_3(\tau,z)^2}{\vartheta_2(\tau,0)^2} \ ,
\end{equation}
and using that the coefficient of the first term is simply the Witten index $\phi_{\rm 1A}(\tau,0)=12$.
On the other hand, this information determines the full elliptic genus via
\begin{equation}\label{rel2}
 \phi_{\rm 1A}(\tau,z)=\phi_{\rm 1A}(\tau,0)\, \frac{\vartheta_2(\tau,z)^2}{\vartheta_2(\tau,0)^2}
+\phi_{\rm 1A}(\tau,\tfrac{1}{2})\, \frac{\vartheta_1(\tau,z)^2}{\vartheta_2(\tau,0)^2} \ .
\end{equation}
Thus the knowledge of $\chi_{\rm 1A}(\tau)$ is equivalent to that of $\phi_{\rm 1A}(\tau,z)$. The
advantage of working with $\chi_{\rm 1A}(\tau)$ is that it is the natural
supersymmetric analogue of the partition function of a bosonic conformal field theory
for which Monstrous Moonshine was originally formulated.

\subsection{Symmetric powers}\label{powers}

For the following it will also be important to consider the symmetric powers of the K3 conformal
field theory. Given any ${\cal N}=2$ superconformal field theory ${\cal H}$, we can consider the
symmetric power theory
\begin{equation}
{\cal H}^{(p)} = \underbrace{\Bigl( {\cal H}{\otimes} \cdots \otimes {\cal H} \Bigr)}_{\hbox{p times}} / {\mathbb Z}_p \ ,
\end{equation}
where the generator $\pi$ of ${\mathbb Z}_p$ is the cyclic permutation of length $p$. We shall concentrate in
the following on the case when $p$ is prime; then all elements of ${\mathbb Z}_p$ are cyclic
permutations of cycle length $p$. Given the elliptic genus of the original ${\cal H}$ theory, we can
then determine the elliptic genus of ${\cal H}^{(p)}$ \cite{DMVV}. From the untwisted sector
of the ${\mathbb Z}_p$ orbifold we get the contribution
\begin{equation}
\phi^{(p)}_{U} (\tau,z) = \frac{1}{p} \Bigl( \bigl(\phi(\tau,z)\bigr)^p + (p-1) \, \phi(p\tau,pz) \Bigr) \ .
\end{equation}
Here the first term is the elliptic genus without the insertion of any orbifold generator, while
the second term gives the contributions coming from $\pi^l$ with $l=1,\ldots, p-1$. Since each
such $\pi^l$ is a cyclic permutation of order $p$, only the states of the form
$(u\otimes u \otimes \cdots \otimes u)$ contribute, and this
leads to $\phi(p\tau,pz)$ for each of these terms.

There are $(p-1)$ twisted sectors, and each of
them contributes the same, thus giving for the whole twisted sector
\begin{equation}
\phi^{(p)}_{T} (\tau,z) =  \frac{(p-1)}{p} \, \sum_{l=0}^{p-1} \phi\bigl(\tfrac{\tau+l}{p},z \bigr) \ ,
\end{equation}
where the sum over $l$ implements the orbifold projection in the twisted sector. Thus
the elliptic genus of the symmetric power theory is (for $p$ prime)
\begin{equation}\label{symorb}
\phi^{(p)}(\tau,z) = \frac{1}{p} \,\bigl( \phi(\tau,z) \bigr)^p + \frac{(p-1)}{p} \, H_p \phi(\tau,z) \ ,
\end{equation}
where $H_p$ denotes the action of the Hecke operator
\begin{equation}\label{Hecke}
H_p \phi (\tau,z) = \phi(p\tau,pz) + \sum_{l=0}^{p-1} \phi\bigl(\tfrac{\tau+l}{p},z \bigr) \ .
\end{equation}
It is well known that (\ref{symorb}) is again a weak Jacobi form of index $mp$
provided that $\phi(\tau,z)$ has index $m$ \cite{DMVV}.
For example, this follows directly from the fact that the Hecke operator
(\ref{Hecke}) maps weak Jacobi forms to weak Jacobi forms, see \cite{EichlerZagier}.

For our purposes, it is also convenient to rewrite  (\ref{Hecke}) for the
NS-sector character $\chi^{(p)}(\tau)$. The result depends on whether $p=2$ or $p$ odd.
For $p=2$ we find
\begin{equation}\label{2}
\chi^{(2)}(\tau) = \chi\left(2\tau,\tfrac{1}{2}\right) + \phi\left(\tfrac{\tau}{2},\tfrac{1}{2}\right) +
(-1)^m \phi\left(\tfrac{\tau}{2}+\tfrac{1}{2},0\right) \ ,
\end{equation}
where $\chi(\tau,z)$ is the NS-sector function defined in (\ref{NS}), while
$\phi(\tau,z)$ is the elliptic genus in the R-sector. In particular, the last expression
is therefore actually a constant since $\phi(\tau,0)$ is the Witten index. The emergence
of the R-sector as the `twisted' sector of the NS-sector is familiar from standard
${\mathbb Z}_2$ orbifold considerations. For the case of the K3 character the relevant formula
is\footnote{We shall state these and any following formulae for the rescaled
elliptic genera,  see (\ref{K3Rr}) and (\ref{K3NS}).}
\begin{eqnarray}
\chi_{\rm 1A}^{(2)}(\tau) & = & \Biggl[ \left(\frac{2 \vartheta_3(2\tau)}{\vartheta_2(2\tau)}\right)^2
-\left(\frac{2 \vartheta_2(2\tau)}{\vartheta_3(2\tau)}\right)^2   \Biggr]
+ \Biggl[
\left(\frac{2 \vartheta_3(\frac{\tau}{2})}{\vartheta_4(\frac{\tau}{2})}\right)^2
-\left(\frac{2 \vartheta_4(\frac{\tau}{2})}{\vartheta_3(\frac{\tau}{2})}\right)^2  \Biggr]
- 12 \nonumber \\
& = & \chi_{\rm 1A}(\tau)^2 + 36 \ .
\end{eqnarray}
The fact that the left-hand side can be expressed in terms of a polynomial of the
original K3 character is a consequence of  $\chi_{\rm 1A}(\tau)^2$ being a Hauptmodul
(without any multiplier system) for the congruence subgroup
$\Gamma_\theta$. Since the left-hand is a modular form for $\Gamma_\theta$ (without
any multiplier system), it must therefore be a polynomial of $\chi_1(\tau)^2$.

For $p>2$ (recall that $p$ is prime, and therefore now odd) the formula is instead
\begin{equation}\label{p}
\chi^{(p)}(\tau) = \chi(p\tau) + \sum_{l=0}^{p-1} \chi\bigl(\tfrac{\tau+4l}{p}\bigr) \ ,
\qquad \hbox{$p\geq 3$ prime.}
\end{equation}
For the case of the K3 character we then find, for example for $p=3$,
\begin{eqnarray}
\chi_{\rm 1A}^{(3)}(\tau) & = & \chi_{\rm 1A}(3\tau) + \chi_{\rm 1A}\left(\tfrac{\tau}{3} \right) +
\chi_{\rm 1A}\left(\tfrac{\tau+4}{3}  \right)  +  \chi_{\rm 1A}\left(\tfrac{\tau+8}{3} \right) \nonumber \\
& = & \chi_{\rm 1A}(\tau)^3 + 60\, \chi_{\rm 1A}(\tau) \ .
\end{eqnarray}
By the same token as above, this formula follows from the fact that the left-hand side
is again a modular form (with multiplier system $-1$ on $S$ and $T^2$) for $\Gamma_\theta$, and
that the Hauptmodul for these is $\chi_{\rm 1A}(\tau)$. Because of the non-trivial multipler system
of the left-hand-side the expression now only involves odd powers of $\chi_{\rm 1A}(\tau)$.

We should note that these relations are the natural analogue of the replication
formulae for the $J$-function, which read for $p=2$ and $p=3$
\begin{eqnarray}\label{Jrep}
J(2\tau) + J\left(\tfrac{\tau}{2}\right) + J\left(\tfrac{\tau+1}{2}\right) & = & J^2(\tau) - 2 a_1 \nonumber \\
J(3\tau) + J\left(\tfrac{\tau}{3}\right) + J\left(\tfrac{\tau+1}{3}\right)
+ J\left(\tfrac{\tau+2}{3}\right) & = & J^3(\tau) - 3 a_1 J(\tau) - 3 a_2 \ .
\end{eqnarray}
Here $J(\tau)$ is the famous $J$-function
\begin{equation}
J(\tau) = \sum_{k=-1}^{\infty} a_k q^k = q^{-1} + 196884 q + 21493760 q^2  + {\cal O}(q^3) \ ,
\end{equation}
that plays a central role in Monstrous moonshine, see \cite{Gannon} for a nice review.

\section{Mathieu symmetry and twining characters}

The elliptic genus of K3 can be thought of as the partition function of the
${\cal N}=2$  half-BPS states of type II  string theory on K3. It is suggested
by the work of  \cite{EOT} that the space of these half-BPS states carries an action of the
Mathieu group ${\mathbb M}_{24}$. Indeed, it was found in \cite{EOT}, following
on from earlier work \cite{EH}, that the elliptic genus of K3 can be written as
\begin{equation}\label{3.1}
\phi_{\rm 1A}(\tau,z) = 24 \, {\rm ch}_{h=\frac{1}{4},l=0}^{\, {\cal N}=4} (\tau,z)
+ \sum_{n=0}^{\infty}  A_n \, {\rm ch}_{h=n+\frac{1}{4},l=\frac{1}{2}}^{\, {\cal N}=4} (\tau,z)\ ,
\end{equation}
where  ${\rm ch}_{h=\frac{1}{4},l=0}^{\, {\cal N}=4}$ is the elliptic genus of the short ${\cal N}=4$
representation with $h=\frac{1}{4}$ and $l=0$ --- see \cite{Eguchi:1987wf,Eguchi:1988af} for
an explicit formula ---  while
\begin{equation}\label{long}
{\rm ch}_{h,l=\frac{1}{2}}^{\, {\cal N}=4} (\tau,z) = q^{h-\frac{3}{8}}\,  \frac{\vartheta_1(\tau,z)^2}{\eta(\tau)^3}
\end{equation}
is the elliptic genus of a long ${\cal N}=4$ representation.\footnote{Strictly speaking, the ${\cal N}=4$
representation with $n=0$ ($h=\tfrac{1}{4}$) is short, and thus (\ref{long}) for $h=\tfrac{1}{4}$ is not
the elliptic genus of a single representation, but rather involves a sum of representations.} The observation
of \cite{EOT} was that the coefficients $A_n$ can be written in terms of dimensions of
irreducible representations of the Mathieu group ${\mathbb M}_{24}$. (The character table of
${\mathbb M}_{24}$, from which one can read off in particular the dimensions of the irreducible
representations is given in the appendix.) More explicitly,
\begin{equation}\label{3.3}
\begin{array}{rclrcl}
24 & = & {\bf 23} + {\bf 1} \qquad \qquad
&  - A_0=2  & = &  {\bf 1} + {\bf 1}  \\
A_1= 90 & = &  {\bf 45} +\overline{\bf 45}  \qquad \qquad
& A_2 = 462 & = & {\bf 231} + \overline{\bf 231} \\
A_3 = 1440 & = & {\bf 770} + \overline{\bf 770} \qquad \qquad
& A_4 = 4554 & = & {\bf 2277} + \overline{\bf 2277} \\
A_5 = 11592 & = & 2\cdot {\bf 5796}  \qquad \qquad
& A_6= 27830 & = & 2\cdot {\bf 3520} + 2\cdot {\bf 10395}
\end{array}
\end{equation}
and
\begin{equation}\label{3.4}
A_7 = 61686 = 2\cdot {\bf 1771} + 2 \cdot {\bf 2024} + 2\cdot {\bf 5313} + 2\cdot {\bf 5796}
+ 2\cdot {\bf 5544} + 2\cdot {\bf 10395} \ .
\end{equation}
Note that, relative to \cite{EOT}, we have absorbed the prefactor of $2$ in \cite[eq.\ (1.11)]{EOT}
into the definition of $A_n$. Then we can write the $A_n$ in terms
of real representations, {\it i.e.}\ sums of representations and their conjugate representations.
Furthermore, for reasons that will be explained momentarily, our expression for $A_7$ differs
from the decomposition suggested in \cite{EOT}. We should mention though that
while $A_7$ given in (\ref{3.4}) is consistent with what is done below, it is not uniquely
specified by this.

\subsection{Explicit formulae for the twining characters}

Obviously, as is apparent from the last comment, there is much arbitrariness in
these decompositions, and they are only a suggestive hint of some underlying
deeper structure. The main aim of this paper is to subject this proposal to a much
more refined test. To this end we shall study the corresponding `{\em twining genus}',
{\it i.e.}\ the analogues of the McKay-Thompson series in the case of Monstrous Moonshine.
This is to say, we shall calculate not  just the elliptic genus itself, but the elliptic genus with the
insertion of a group element $g\in {\mathbb M}_{24}$
\begin{equation}
\phi_g(\tau,z) = \hbox{Tr}_{{\cal H}_{RR}}\Bigl( g\, q^{L_0-\frac{c}{24}} e^{2\pi i z J_0} (-1)^F \,
\bar{q}^{\bar{L}_0-\frac{\bar{c}}{24}} (-1)^{\bar{F}} \Bigr) \ .
\end{equation}
Technically speaking, the calculation of $\phi_g$ differs from $\phi$  by
replacing $A_n$ in (\ref{3.1}) by  the characters of $g$ in the representations
described in (\ref{3.3}). Obviously, the character of $g$ only depends  on its
conjugacy class, and thus all the information we need is contained in the character
table of  ${\mathbb M}_{24}$ that is included in the appendix.

One immediate problem with this approach is that we only know how to write
$A_n$ in terms of dimensions of representations of ${\mathbb M}_{24}$  for the first few values of $n$.
We can therefore only directly determine the low powers of $q$ of the twining
genera $\phi_g$. If this was all there was to it, this would not be particularly
interesting, and it would certainly not lead to any non-trivial consistency check of the
proposal. The reason why interesting information can be obtained in this manner
is that the twining genera must have fairly specific modular properties. One way
to see this explicitly is to translate the R-sector series $\phi_g(\tau,z)$ into the
NS-sector as in (\ref{NS}), and set $z=0$. The resulting expression is the
NS-sector twining character $\chi_g(\tau)$, {\it i.e.}\ the character in the NS-sector
with the insertion of the Mathieu group element $g$. Note that by the same arguments as
above  we are not loosing any information in doing so, because the analogues of (\ref{rel1}) and
(\ref{rel2}) are
\begin{equation}\label{rel1g}
 \chi_g(\tau,z) = \phi_g(\tau,0)\, \frac{\vartheta_4(\tau,z)^2}{\vartheta_2(\tau,0)^2}
-\phi_g(\tau,\tfrac{1}{2})\frac{\vartheta_3(\tau,z)^2}{\vartheta_2(\tau,0)^2} \ ,
\end{equation}
and
\begin{equation}\label{rel2g}
 \phi_g(\tau,z)=\phi_g(\tau,0)\, \frac{\vartheta_2(\tau,z)^2}{\vartheta_2(\tau,0)^2}
+\phi_g(\tau,\tfrac{1}{2})\, \frac{\vartheta_1(\tau,z)^2}{\vartheta_2(\tau,0)^2} \ ,
\end{equation}
where now the constant term equals
\begin{equation}
\phi_g(\tau,0)= \tfrac{1}{2}\, \Tr_{{\bf 23}\oplus{\bf 1}}(g) \ .
\end{equation}
Thus we can reconstruct the full twining genus $\phi_g(\tau,z)$ from the twining character
$\chi_g(\tau)$, and vice versa. In terms of the usual box notation, $\chi_g(\tau)$
corresponds to
\begin{eqnarray}\label{boxd}
\chi_g(\tau) \longleftrightarrow
& (-1,{\bf 1})\, &   \boxed{\phantom{{ \tiny \begin{array}{cc}
				a & aa \\a & a \end{array}}}} \nonumber \\
& &  \hspace*{-0.1cm} (-1,{\bf g})
\end{eqnarray}
where the '$-1$'s indicate that the trace is taken in the NS-sector,
and without any insertion of $(-1)^F$, respectively. Under a
modular transformation it is believed that these twining and
twisted characters transform as
\begin{equation}
\begin{array}{llllll}
& h & \boxed{\phantom{{\tiny \begin{array}{cc}
				a & aa \\a & a \end{array}}}}
\hspace*{0.5cm}
& \stackrel{\left(\begin{array}{cc} a & b \\ c & d \end{array}\right)} {\xrightarrow{\hspace*{2cm}}} \hspace*{0.5cm}
& h^d g^c & \boxed{\phantom{{\tiny \begin{array}{cc}
				a & aa \\a & a \end{array}}}}  \\
& & \hspace*{0.3cm}  g
& & &  \hspace*{0.1cm} g^a h^b
\end{array}
\end{equation}
For the case at hand, it therefore follows that the twining character
corresponding to (\ref{boxd}) is invariant (possibly up to some phase)
under a modular transformation provided that
\begin{equation}\label{cons}
a+b, \ c+d  \in 2 {\mathbb Z} +1\ , \qquad c = 0 \quad \hbox{mod $o(g)$,}
\end{equation}
where $o(g)$ is the order of the group element $g$. [Here we have used that the
group operation is $(\epsilon_1,g_1) \cdot (\epsilon_2,g_2) = (\epsilon_1\epsilon_2, g_1 \cdot g_2)$,
where $\epsilon_{j}\in\{\pm 1\}$ and $g_i\in {\mathbb M}_{24}$. We have furthermore used
that the conditions in (\ref{cons}) together with the requirement that $ad-bc=1$ imply
that $(a,o(g))=1$. For the Mathieu group ${\mathbb M}_{24}$ it then follows that $g^a$ is in
the same conjugacy class as $g$.]
The subgroup of $SL(2,{\mathbb Z})$ that is characterised by (\ref{cons}) contains in particular
\begin{equation}\label{3.5}
\Gamma^g = \Gamma_\theta \cap \Gamma_0(o(g)) \ ,
\end{equation}
where $\Gamma_\theta$ was introduced before, and
\begin{equation}
\Gamma_0(N) = \left\{
\left( \begin{array}{cc} a & b \\ c & d \end{array} \right) \in
SL(2,{\mathbb Z}) \ : \  c = 0 \,\, \hbox{mod $(N)$} \right\}  \ .
\end{equation}
The property of a function to be modular invariant under $\Gamma^g$
(possibly up to some multiplier system) is a very strong condition, and knowing
the first few terms usually determines the function uniquely.

Thus the non-trivial consistency test we will be performing proceeds as follows. Using
the low-lying decomposition (\ref{3.3}), we can calculate the first few coefficients of the
twining characters $\chi_g(\tau)$ explicitly. On the other hand, we know that these functions
must be invariant under $\Gamma^g$, and
thus there is the highly non-trivial consistency condition whether this ansatz can be
completed in a $\Gamma^g$-invariant manner. The results we find are consistent
with $A_n$ given in (\ref{3.3}) and (\ref{3.4}), but not with the formula for $A_7$ suggested in
\cite{EOT}. They represent a highly non-trivial consistency condition for the conjecture
of \cite{EOT} that the elliptic genus of K3 carries an action by the Mathieu group ${\mathbb M}_{24}$.

The constraints coming from the modular properties are tighter the larger $\Gamma^g$ is,
{\it i.e.}\ the smaller the order of the group element is. In
particular, we should therefore expect to have good control over the twining characters
corresponding to group elements of small order. Indeed, this turns out to be correct, and we have
even found closed form expressions that match with the low lying coefficients and have the correct
modular properties.  More specifically, we find that --- in the following the label refers to the conjugacy
class of the Mathieu group element that is described in the appendix
\begin{align}
& \chi_{\rm 1A}(\tau)  =  \kappa(\tau)
=T_{\rm [8a]}(\tfrac{\tau}{4})  \qquad  \hbox{ --- see   eq.\ (\ref{K3NS})} \nonumber  \\[4pt]
& \chi_{{\rm 2A}}(\tau)  =  {\displaystyle 4\,   \frac{\vartheta_4(\tau)^2}{\vartheta_2(\tau)^2} }
 \simeq T_{\rm [16b]}(\tfrac{\tau}{4})  \nonumber  \\[4pt]
& \chi_{{\rm 2B}}(\tau) =  {\displaystyle 4\,  \frac{\vartheta_3(\tau)^2}{\vartheta_2(\tau)^2} }
\simeq T_{\rm [16b]}(\tfrac{\tau}{4})  \nonumber  \\[6pt]
& \chi_{{\rm 3A}}(\tau) =  {\displaystyle \left( \frac{\eta(\tau)}{\eta(3\tau)} \right)^3\,
\frac{\vartheta_3(3\tau)}{\vartheta_3(\tau)}
=T_{\rm [24c]}(\tfrac{\tau}{4})}  \nonumber \\[6pt]
&\chi_{\rm 4A}(\tau) =  {\displaystyle \left(\frac{\eta(\tau) \,
\eta(2\tau)}{\eta(\frac{\tau}{2}) \, \eta(4\tau)}\right)^4}  =T_{\rm [16A]}(\tfrac{\tau}{4}) \label{3.11} \\[4pt]
& \chi_{\rm 4B}(\tau) =  {\displaystyle 2\, \frac{\vartheta_3(2\tau)}{\vartheta_2(2\tau)}}
 =T_{\rm [16B]}(\tfrac{\tau}{4}) \nonumber  \\[6pt]
&\chi_{\rm 4C}(\tau) =  {\displaystyle 4\, \frac{\vartheta_3(\tau)^2 \vartheta_3(2\tau)}
{\vartheta_2(\tau)^2 \vartheta_4(2\tau)}}
\nonumber \\[6pt]
&\chi_{\rm 5A}(\tau)  =
q^{-\frac{1}{4}} \Bigl( 1 + 3\, q + 4\, q^{\frac{3}{2}} + 4 \, q^2 + 4\, q^{\frac{5}{2}} + \cdots \Bigr)
 = T_{\rm [40a]}(\tfrac{\tau}{4})   \nonumber \\[6pt]
&\chi_{\rm 6A}(\tau)  =  {\displaystyle \left(\frac{\eta(\frac{3}{2}\tau) \,
\eta(2\tau)}{\eta(\frac{\tau}{2}) \, \eta(6\tau)}\right)^2}  = T_{\rm [24H]}(\tfrac{\tau}{4}) \ . \nonumber
\end{align}
Our conventions for the Jacobi theta functions, and the first few
coefficients of the power series expansions of the relevant McKay-Thompson series
are given in the appendix. We have again
normalised these functions so that they start with $1\cdot q^{-\frac{1}{4}}$,
compare (\ref{K3NS}). In most cases, the functions can be identified with a standard McKay-Thompson
series $T_{\rm [*]}$ (evaluated at $\frac{\tau}{4}$) --- if this is the case the corresponding
identity is given above. For the  conjugacy class 2A and 2B the
coefficients only agree with the McKay-Thompson series up to some signs; this is indicated by
$\simeq$. In all cases we have explicitly checked that these functions are
invariant under $\Gamma^g$ in (\ref{3.5}) up to some multiplier system. The multiplier
system is non-trivial in all cases since all twining characters transform as
$-1$ under $T^2\in \Gamma^g$. In addition, $\chi_{\rm 1A}$ transforms as $-1$ under $S$
(compare (\ref{eq:NSegtmrn})), while $\chi_{\rm 4C}$ transforms as
$-i$ under $S T^4 S$. (We have not checked the precise form of the multiplier system for the
class corresponding to 5A.)

The attentive reader will notice that (\ref{3.11}) accounts for all twining characters of group elements
up to order $6$, except for the ones corresponding to 3B and 6B. While we have not managed to find
any closed formulae for them --- explicit formulae for the first few coefficients of their
power series expansion are given in the appendix --- we have an indirect argument that
suggests that at least the twining character corresponding to 3B has good modular properties;
this will be explained in the following section. We should also mention that some of these conjugacy
classes are actually contained in ${\mathbb M}_{23}$ --- these are the conjugacy classes
1A, 2A, 3A, 4B, 5A, 6A --- while the others (2B, 3B, 4A, 4C) lie only in ${\mathbb M}_{24}$ but
not in ${\mathbb M}_{23}$. These latter classes are characterised by the property that
$\Tr_{{\bf 23}\oplus{\bf 1}}(g)=0$. For them the relation between the twining characters
and the twining genera (see (\ref{rel1g}) and (\ref{rel2g})) is therefore particularly simple, and we
have
\begin{equation}
\phi_g(\tau,z) = - \chi_g(\tau) \, \frac{\vartheta_1(\tau,z)^2}{\vartheta_3(\tau,0)^2} \qquad \qquad
\hbox{if $\Tr_{{\bf 23}\oplus{\bf 1}}(g)=0$.}
\end{equation}
Thus for the elements in ${\mathbb M}_{24}$ that are not in ${\mathbb M}_{23}$, the
modular properties of $\phi_g(\tau,z)$ are  a direct consequence of those
of $\chi_g(\tau)$. However, apart from this simplification, we do not see any
difference between the behaviour of the elements that lie in ${\mathbb M}_{23}$
and those that do not.

\subsection{Replication formulae}\label{sec:3.1}

One of the remarkable properties of the McKay-Thompson series in Monstrous Moonshine
are their replication relations. They are a direct generalisation of the $J$-function replication
identities (\ref{Jrep}) to the McKay-Thompson series. In the following we want to give
a conformal field theoretic argument that suggests that similar relations should also
hold for the twining characters of K3, and more specifically, what precise form they should
take. Incidentally, if we applied the following argument to the Monster theory it would give a physics
explanation of the replication identities of Monstrous Moonshine.

The idea of the argument is very simple. If the ${\cal N}=2$ half-BPS spectrum of K3 carries an action of the
Mathieu group ${\mathbb M}_{24}$, then the same will apply to the symmetric power theories
where $g\in {\mathbb M}_{24}$ acts diagonally on all $p$ factors so as to commute with
the orbifold action. But if this is so, then we can also consider the twining character
corresponding to $g\in {\mathbb M}_{24}$ for the $p^{\rm th}$ symmetric power theory.
Given our detailed understanding of the orbifold theory, we can determine how
$g$ acts on the various terms in (\ref{symorb}) or indeed on the different terms in
$H_p \phi(\tau,z)$.  The first term corresponds to the untwisted sector contribution
where the cyclic permutation generator has been inserted into the trace. This implies that
only the states of the form $(u\otimes u \otimes \cdots \otimes u)$ contribute. Inserting
the diagonal group element $g\in {\mathbb M}_{24}$ then leads to $\phi_{g^p}(p\tau,pz)$.
The other terms of $H_p \phi(\tau,z)$ come from the twisted sector, and the natural
action of  $g\in {\mathbb M}_{24}$ is simply described by replacing $\phi$ by $\phi_g$. Thus
we are led to claim that the twining genus for the Hecke part of the
$p^{\rm th}$ symmetric power theory is given by
\begin{equation}
\bigl( H_p \phi\bigr)_g (\tau,z) = \phi_{g^p}(p\tau,pz) + \sum_{l=0}^{p-1} \phi_g\bigl(\tfrac{\tau+l}{p},z \bigr) \ .
\end{equation}
Translated into the NS-sector characters we then have for $p=2$
\begin{equation}\label{2p}
\chi_g^{(2)}(\tau) = \chi_{g^2}\left(2\tau,\tfrac{1}{2}\right) + \phi_g\left(\tfrac{\tau}{2},\tfrac{1}{2}\right) +
(-1)^m \phi_g\left(\tfrac{\tau}{2}+\tfrac{1}{2},0\right) \ ,
\end{equation}
while for  $p>2$ the formula is
\begin{equation}\label{pp}
\chi^{(p)}_g(\tau) = \chi_{g^p}(p\tau) + \sum_{l=0}^{p-1}
\, \chi_g\left(\tfrac{\tau+4l}{p} \right) \ ,
\qquad \hbox{$p\geq 3$ prime.}
\end{equation}
On the other hand, it is clear from the discussion around (\ref{3.5}) that also these
twining characters must be invariant under $\Gamma^g$, possibly up to a (different) multiplier
system. In the context of Monstrous Moonshine, the McKay-Thompson series were Hauptmoduls
for the corresponding genus zero congruence subgroups, and thus the McKay-Thompson series
of the symmetric power theories could be written as a polynomial of the original McKay-Thompson
series. This then leads to the replication formulae of Monstrous Moonshine (generalising
directly the $J$-function replication identities of (\ref{Jrep})).

In the current context, it appears that the twining characters are {\em not} of genus zero,
but we still find remarkable replication identities that give yet another consistency check
on our analysis (and indicate that the twining characters corresponding to 3B and 7A=7B
have indeed good modular properties). Specifically, we have found the following identities for
$p=2$
\begin{align}
&\chi_{\rm 2A}^{(2)}(\tau) = \chi_{\rm 1A}\left(2\tau,\tfrac{1}{2}\right)
+\phi_{\rm 2A}\left(\tfrac{\tau}{2},\tfrac{1}{2}\right) - 4
=\chi_{\rm 2A}(\tau)^2+4  \\[4pt]
&\chi_{\rm 2B}^{(2)}(\tau) = \chi_{\rm 1A}\left(2\tau,\tfrac{1}{2}\right)
+\phi_{\rm 2B}\left(\tfrac{\tau}{2},\tfrac{1}{2}\right)
=\chi_{\rm 2B}(\tau)^2-12 \\[4pt]
&\chi_{\rm 3A}^{(2)}(\tau) = \chi_{\rm 3A}\left(2\tau,\tfrac{1}{2}\right)
+\phi_{\rm 3A}\left(\tfrac{\tau}{2},\tfrac{1}{2}\right) - 3
=\chi_{\rm 3A}(\tau)^2 \\[4pt]
&\chi_{\rm 3B}^{(2)}(\tau) = \chi_{\rm 3B}\left(2\tau,\tfrac{1}{2}\right)
+\phi_{\rm 3B}\left(\tfrac{\tau}{2},\tfrac{1}{2}\right)=-\tfrac{1}{2}\,\chi_{\rm 3B}(\tau)^2
+\tfrac{3}{2}\,T_{\rm [12D]}\left(\tfrac{\tau}{2}\right)  \label{3Bp=2}\\[4pt]
&\chi_{\rm 4A}^{(2)}(\tau) =\chi_{\rm 2A}\left(2\tau,\tfrac{1}{2}\right)
+\phi_{\rm 4A}\left(\tfrac{\tau}{2},\tfrac{1}{2}\right)=-\chi_{\rm 4A}(\tau)^2
+2\,T_{\rm [8E]}\left(\tfrac{\tau}{2}\right)+4\\[4pt]
&\chi_{\rm 4B}^{(2)}(\tau) = \chi_{\rm 2A}\left(2\tau,\tfrac{1}{2}\right)
+\phi_{\rm 4B}\left(\tfrac{\tau}{2},\tfrac{1}{2}\right) - 2
=- \chi_{\rm 4B}(\tau)^2+2\,T_{\rm [8E]}\left(\tfrac{\tau}{2}\right)-4 \\[4pt]
&\chi_{\rm 4C}^{(2)}(\tau) =  \chi_{\rm 2B}\left(2\tau,\tfrac{1}{2}\right)
+\phi_{\rm 4C}\left(\tfrac{\tau}{2},\tfrac{1}{2}\right)
=-\tfrac{1}{2}\,\chi_{\rm 4C}(\tau)^2+\tfrac{3}{2}\,T_{\rm [16d]}\left(\tfrac{\tau}{4}\right)^2 \\[4pt]
&\chi_{\rm 5A}^{(2)}(\tau) = \chi_{\rm 5A}\left(2\tau,\tfrac{1}{2}\right)
+\phi_{\rm 5A}\left(\tfrac{\tau}{2},\tfrac{1}{2}\right) - 2
=-\chi_{\rm 5A}(\tau)^2+2\,T_{\rm [20C]}\left(\tfrac{\tau}{2}\right)-4 \\[4pt]
&\chi_{\rm 6A}^{(2)}(\tau) = \chi_{\rm 3A}\left(2\tau,\tfrac{1}{2}\right)
+\phi_{\rm 6A}\left(\tfrac{\tau}{2},\tfrac{1}{2}\right) -1 =
-\chi_{\rm 6A}(\tau)^2+2 \,T_{\rm [12I]}\left(\tfrac{\tau}{2}\right) \\[4pt]
&\chi_{\rm 7A}^{(2)}(\tau) = \chi_{\rm 7A}\left(2\tau,\tfrac{1}{2}\right)
+\phi_{\rm 7A}\left(\tfrac{\tau}{2},\tfrac{1}{2}\right) - \tfrac{3}{2}
=- 2 \chi_{\rm 7A}(\tau)^2+3\, T_{\rm [28B]}\left(\tfrac{\tau}{2}\right) \ .
\end{align}
These identities imply recursion relations
for the twining characters which one could use, say, to determine the twining character for 3B explicitly.
Furthermore, since the McKay-Thompson series are known to have good modular
transformation properties, the existence of these relations implies that the same
must be true for the twining characters that appear here.

\noindent For $p=3$ we have found the following explicit identities
\begin{align}
& {\displaystyle \chi_{\rm 2A}^{(3)}(\tau) =  \chi_{\rm 2A}(3\tau)
+ \sum_{l=0}^2\chi_{\rm 2A}\left(\tfrac{\tau+4l}{3}\right)
=\chi_{\rm 2A}(\tau)^3+12\,\chi_{\rm 2A}(\tau)} \\[2pt]
& {\displaystyle \chi_{\rm 2B}^{(3)}(\tau) =
\chi_{\rm 2B}(3\tau)+\sum_{l=0}^2\chi_{\rm 2B}\left(\tfrac{\tau+4l}{3}\right)
=\chi_{\rm 2B}(\tau)^3-12\,\chi_{\rm 2B}(\tau) }\\[2pt]
& {\displaystyle \chi_{\rm 3A}^{(3)}(\tau) =
\chi_{\rm 1A}(3\tau)+\sum_{l=0}^2\chi_{\rm 3A}\left(\tfrac{\tau+4l}{3}\right)
=\chi_{\rm 3A}(\tau)^3+6\,\chi_{\rm 3A}(\tau)} \\[2pt]
& {\displaystyle \chi_{\rm 4A}^{(3)}(\tau) =
\chi_{\rm 4A}(3\tau)+\sum_{l=0}^2\chi_{\rm 4A}\left(\tfrac{\tau+4l}{3}\right)
=\chi_{\rm 4A}(\tau)^3-12\,\chi_{\rm 4A}(\tau) }\\[2pt]
& {\displaystyle \chi_{\rm 4B}^{(3)}(\tau) =
\chi_{\rm 4B}(3\tau)+\sum_{l=0}^2\chi_{\rm 4B}\left(\tfrac{\tau+4l}{3}\right)
=\chi_{\rm 4B}(\tau)^3} \\[2pt]
& {\displaystyle \chi_{\rm 4C}^{(3)}(\tau) =
\chi_{\rm 4C}(3\tau)+\sum_{l=0}^2\chi_{\rm 4C}\left(\tfrac{\tau+4l}{3}\right)=
\tfrac{1}{4}\chi_{\rm 4C}(\tau)^3+\tfrac{3}{4}\chi_{\rm 4C}(\tau)\,
T_{\rm [16d]}\left(\tfrac{\tau}{4}\right)^2-6\,T_{\rm [16d]}\left(\tfrac{\tau}{4}\right)} \\
& {\displaystyle \chi_{\rm 5A}^{(3)}(\tau) =
\chi_{\rm 5A}(3\tau)+\sum_{l=0}^2\chi_{\rm 5A}\left(\tfrac{\tau+4l}{3}\right)
=\chi_{\rm 5A}(\tau)^3 }\\[2pt]
& {\displaystyle \chi_{\rm 6A}^{(3)}(\tau) =
\chi_{\rm 2A}(3\tau)+\sum_{l=0}^2\chi_{\rm 6A}\left(\tfrac{\tau+4l}{3}\right)
=\chi_{\rm 6A}(\tau)^3-6\,\chi_{\rm 6A}(\tau) \ ,}
\end{align}
while for $p=5$ we have found
\begin{align}
&{\displaystyle
\chi_{\rm 2A}^{(5)}(\tau) =
\chi_{\rm 2A}(5\tau)+\sum_{l=0}^4\chi_{\rm2A}\left(\tfrac{\tau+4l}{5}\right)
= \chi_{\rm 2A}(\tau)^5+20\,\chi_{\rm 2A}(\tau)^3+70\,\chi_{\rm 2A}(\tau) }\\[2pt]
&{\displaystyle
\chi_{\rm 2B}^{(5)}(\tau) =
\chi_{\rm 2B}(5\tau)+\sum_{l=0}^4\chi_{\rm 2B}\left(\tfrac{\tau+4l}{5}\right)
=\chi_{\rm 2B}(\tau)^5-20\,\chi_{\rm 2B}(\tau)^3+70\,\chi_{\rm 2B}(\tau) } \\[2pt]
&{\displaystyle
\chi_{\rm 3A}^{(5)}(\tau) =
\chi_{\rm 3A}(5\tau)+\sum_{l=0}^4\chi_{\rm 3A}\left(\tfrac{\tau+4l}{5}\right)
=\chi_{\rm 3A}(\tau)^5+10\,\chi_{\rm 3A}(\tau)^3+15\,\chi_{\rm 3A}(\tau)} \\[2pt]
&{\displaystyle
\chi_{\rm 4A}^{(5)}(\tau) =
\chi_{\rm 4A}(5\tau)+\sum_{l=0}^4\chi_{\rm 4A}\left(\tfrac{\tau+4l}{5}\right)
=\chi_{\rm 4A}(\tau)^5-20\,\chi_{\rm 4A}(\tau)^3+30\,\chi_{\rm 4A}(\tau)} \\[2pt]
&{\displaystyle
\chi_{\rm 4B}^{(5)}(\tau) =
\chi_{\rm 4B}(5\tau)+\sum_{l=0}^4\chi_{\rm 4B}\left(\tfrac{\tau+4l}{5}\right)
= \chi_{\rm 4B}(\tau)^5-10\,\chi_{\rm 4B}(\tau)} \\[2pt]
&{\displaystyle
\chi_{\rm 4C}^{(5)}(\tau) =
\chi_{\rm 4C}(5\tau)+\sum_{l=0}^4\chi_{\rm 4C}\left(\tfrac{\tau+4l}{5}\right)
=\tfrac{1}{16}\,\chi_{\rm 4C}(\tau)^5
-\tfrac{1}{16}\chi_{\rm 4C}(\tau)^3\,T_{\rm [16d]}\left(\tfrac{\tau}{4}\right)^2
} \nonumber\\
&\qquad \qquad\qquad
{\displaystyle
+\chi_{\rm 4C}(\tau)\,T_{\rm [16d]}\left(\tfrac{\tau}{4}\right)^4
-\tfrac{1}{2}\,\chi_{\rm 4C}(\tau)^2\,T_{\rm [16d]}\left(\tfrac{\tau}{4}\right)
-4\,T_{\rm [16d]}\left(\tfrac{\tau}{4}\right)^3
-2\,\chi_{\rm 4C}(\tau)} \\[2pt]
&{\displaystyle
\chi_{\rm 5A}^{(5)}(\tau) =
\chi_{\rm 1A}(5\tau)+\sum_{l=0}^4\chi_{\rm 5A}\left(\tfrac{\tau+4l}{5}\right)
=\chi_{\rm 5A}(\tau)^5-15\,\chi_{\rm 5A}(\tau)} \\[2pt]
&{\displaystyle
\chi_{\rm 6A}^{(5)}(\tau) =
\chi_{\rm 6A}(5\tau)+\sum_{l=0}^4\chi_{\rm 6A}\left(\tfrac{\tau+4l}{5}\right)
=\chi_{\rm 6A}(\tau)^5-10\,\chi_{\rm 6A}(\tau)^3-5\,\chi_{\rm 6A}(\tau) \ . }
\end{align}
These are probably not the only identities that exist, but only those
we have been able to find with our current partial understanding of the twining
characters. Indeed, in order to be able to determine the higher order  replication formulae,
one needs to know the twining characters up to some fairly high level. Thus
with our current understanding we can only check these identities for the twining
characters given in (\ref{3.11}).

%%%%%%%%%%%%%%%%%%%%%%%%%%%%%%%%%%%%%%%%%%

\section{Conclusions}

In this paper we have provided strong evidence for the conjecture of
Eguchi, Ooguri and Tachikawa \cite{EOT} that the elliptic genus of K3 carries
a natural action of the Mathieu group ${\mathbb M}_{24}$. In particular, we
have shown that the elliptic genus with the insertion of a group element
in ${\mathbb M}_{24}$ leads, in the NS-sector, to a twining character that possesses
nice modular properties. By considering the twining characters of symmetric power
theories we have furthermore shown that the twining characters satisfy beautiful
replication identities similar to those appearing for the McKay-Thompson series
in Monstrous Moonshine. While the analogy of the twining characters with the
McKay-Thompson series is quite striking, there are also marked differences:
in the current context the twining characters are not Hauptmoduls for a genus
zero congruence subgroup, and hence the replication relations contain
also `inhomogeneous' terms (for example the term proportional to $T_{\rm [12D]}$
in (\ref{3Bp=2})).
\smallskip

So far we have only worked out about half of the twining characters in detail.
In order to get a more complete understanding of the Mathieu action on the
K3 elliptic genus, it would be very desirable to find closed form expressions
for all twining characters. Probably some more refined techniques (rather than
just guessing and confirming the answer) will be necessary; in particular,
the techniques developed by Bantay \& Gannon \cite{BG1,BG2} should be useful for
this.
\smallskip

The above analysis suggests that the space of ${\cal N}=2$ half-BPS states
of K3 carries an action of the Mathieu group ${\mathbb M}_{24}$.  It is believed that
these ${\cal N}=2$ half-BPS states form a subset of the
quarter-BPS states (with respect to ${\cal N}=4$), for which the spectrum generating algebra
is the so-called BPS-algebra of Harvey \& Moore \cite{HM,DVV}.
The above group action on the spectrum then suggests that
the Mathieu group ${\mathbb M}_{24}$ should be (contained in) the
automorphism group of that Borcherds algebra, in complete analogy to what happened
for Monstrous Moonshine. It would be very interesting to explore
this idea in more detail.

\section*{Acknowledgments}
We thank Terry Gannon for useful communications and Daniel Persson for
comments on the draft. The research of MRG is partially
supported by a grant from the Swiss National Science Foundation, and the research of RV
is supported by an INFN Fellowship.
\bigskip

\noindent{\it Note added:} As this paper was being finalised, a preprint
appeared {\tt arXiv:1005.5415} \cite{Cheng:2010pq} that contains overlapping results
with those of section 3.1 of the present paper.

%%%%%%%%%%%%%%%%%%%%%%%%%%%%%%%%%%%%%%%%%%%%%%%%%%%%%%%%%%%%%%%%%%%%%%%%%%%

\appendix

\section{Definitions}
Our conventions for  the Dedekind eta and the
Jacobi theta functions are
\begin{align}
\eta(\tau) & =  q^{\frac{1}{24}} \prod_{n=1}^{\infty} (1 - q^n) \nonumber \\
\vartheta_1(\tau,z) & = -iq^{\frac{1}{8}} y^{\frac{1}{2}}\, \prod_{n=1}^\infty(1-q^n)(1-yq^n)(1-y^{-1}q^{n-1})
\nonumber \\
\vartheta_2(\tau,z) & = 2\, q^{\frac{1}{8}} \cos(\pi z)\, \prod_{n=1}^{\infty} (1-q^n)\, (1+yq^n)
(1+y^{-1} q^n)   \nonumber \\
\vartheta_3(\tau,z) & =    \prod_{n=1}^{\infty} (1-q^n) \, (1+yq^{n-1/2})(1+y^{-1}q^{n-1/2}) \\
\vartheta_4(\tau,z) & =    \prod_{n=1}^{\infty} (1-q^n) \, (1-yq^{n-1/2}) (1-y^{-1}q^{n-1/2}) \ \ .\nonumber
\end{align}
Under modular transformations the $\vartheta$  and $\eta$ functions transform as
\begin{align}
&\eta(\tau+1) =  e^{ \frac{2\pi i}{24}} \, \eta(\tau) \quad
&\vartheta_2(\tau+1,z) = e^{ \frac{2\pi i}{8}} \, \vartheta_2(\tau,z)  \ , \\
&\vartheta_3(\tau+1,z) = \vartheta_4(\tau,z)  \quad
&\vartheta_4(\tau+1,z) = \vartheta_3(\tau,z) \ ,
\end{align}
as well as
\begin{align}
&\eta(-\tfrac{1}{\tau}) = (-i \tau)^{\frac{1}{2}}\, \eta(\tau)  \quad
&\vartheta_2(- \tfrac{1}{\tau},\tfrac{z}{\tau}) = (-i \tau)^{\frac{1}{2}}\,
e^{\frac{i\pi z^2}{\tau}}\,  \vartheta_4(\tau,z) \ , \\
&\vartheta_3(- \tfrac{1}{\tau},\tfrac{z}{\tau}) =(-i \tau)^{\frac{1}{2}}\,
e^{\frac{i\pi z^2}{\tau}}\,  \vartheta_3(\tau,z)  \quad
&\vartheta_4(- \tfrac{1}{\tau},\tfrac{z}{\tau}) = (-i \tau)^{\frac{1}{2}}\,
e^{\frac{i\pi z^2}{\tau}}\,  \vartheta_2(\tau,z) \ .
\end{align}
The theta functions $\vartheta_a(\tau)$ are defined as $\vartheta_a(\tau)\equiv \vartheta_a(\tau,z=0)$.

\subsection{McKay-Thompson series}

For completeness we give here the first few terms of the McKay-Thompson series that
appear in our analysis
\begin{align}
&T_{\rm [8D]}=q^{-1}+4q+2q^3+8q^5-q^7+\cdots\\
&T_{\rm [8E]}=q^{-1}+4q+2q^3-8q^5-q^7+\cdots\\
&T_{\rm [8a]} = q^{-1} - 20q -62 q^3 - 216 q^5 -641 q^7 - 1636 q^9 -  \cdots \\
&T_{\rm [12D]}=q^{-1}+8q^2+28q^5+ 64 q^8 +134 q^{11} + \cdots \\
&T_{\rm [12I]}=q^{-1}+2q+q^3 - 2 q^7 - 2q^9 +2 q^{11} + \cdots \\
&T_{\rm [16A]}=q^{-1} + 4q +  10 q^3+ 24 q^5 + 47 q^7 + 84q^9 + \cdots \\
&T_{\rm [16B]}=q^{-1}  + 2q^3  - q^7 - 2 q^{11} + 3 q^{15} + 2 q^{19} + \cdots\\
&T_{\rm [16b]}=q^{-1} + 4q  - 2q^3 +8 q^5 - q^7 +20 q^9 + \cdots \\
&T_{\rm [16d]}=T_{\rm [8D]}(2\tau)^{1/2}=q^{-1} - 2q^3 - q^7 + 2q^{11} +\cdots \\
&T_{\rm [20C]}=q^{-1} +q-2q^2+2q^3+2q^4-q^5-4q^7+\cdots \\
&T_{\rm [24H]}=q^{-1} +  2q + 5 q^3 + 8 q^5 +14 q^7 + 22 q^9 + \cdots\\
&T_{\rm [24c]}=q^{-1} - 2 q + q^3 - 2 q^7 + 2 q^9 + 2 q^{11} + \cdots\\
&T_{\rm [28B]}=q^{-1}+3q+4q^2+9q^3+12q^4+15q^5+24q^6+39q^7+\cdots \\
&T_{\rm [40a]}=q^{-1}+ 3 q^3 + 4 q^{5} + 4q^7 + 4q^{9} + 7 q^{11} + \cdots \ ,
\end{align}
see  \cite{FMN} and {\tt http://www.research.att.com/~njas/sequences/} for more information about these
series.

\subsection{Twining characters for K3}

Here we give the low order expansion of the remaining twining characters. Since the spectrum
of the BPS states is real, the twining characters of conjugacy classes that only differ on
conjugate representations agree. The remaining functions are therefore
\begin{align}
&\chi_{\rm 3B}(\tau) = q^{-\frac{1}{4}} \Bigl( 1 + 4 q^{\frac{1}{2}} + 4 q + 4 q^{2} + 8 q^{\frac{5}{2}} + 2 q^3
            + 8 q^{\frac{7}{2}} + 12 q^4 + 4 q^{\frac{9}{2}} + 16 q^5 + 16 q^{\frac{11}{2}} \nonumber \\
&  \qquad \qquad  \qquad        + 5 q^6 +26 q^{\frac{13}{2}} + 21 q^7 + \cdots \Bigr) \\[4pt]
&\chi_{\rm 6B}(\tau) = q^{-\frac{1}{4}} \Bigl( 1+ 4 q^{\frac{1}{2}} + 8 q + 16 q^{\frac{3}{2}}
            + 32 q^{2} + 56 q^{\frac{5}{2}} + 94 q^3  +
            + 152 q^{\frac{7}{2}} + 240 q^4 + 372 q^{\frac{9}{2}} \nonumber \\
 &  \qquad \qquad  \qquad         + 560 q^5 + 832 q^{\frac{11}{2}}
            + 1197 q^6 +1594 q^{\frac{13}{2}} + 1849 q^7  + \cdots \Bigr)  \\[4pt]
&\chi_{\rm 7A}(\tau) = \chi_{\rm 7B}(\tau) = q^{-\frac{1}{4}} \Bigl( 1 + q^{\frac{1}{2}} + \tfrac{9}{2} q
            + 8 q^{\frac{3}{2}} + \tfrac{27}{2} q^{2} +  23 q^{\frac{5}{2}} + \tfrac{81}{2} q^3
            + 68 q^{\frac{7}{2}} + 104 q^4 \nonumber \\
&  \qquad \qquad  \qquad   \qquad
+ 154 q^{\frac{9}{2}} + \tfrac{469}{2}q^5 + 352 q^{\frac{11}{2}}
+ \tfrac{987}{2} q^6 + \tfrac{1293}{2}q^{\frac{13}{2}} + 751 q^7 + \cdots \Bigr) \\[4pt]
&\chi_{\rm 8A}(\tau) = q^{-\frac{1}{4}} \Bigl( 1 + 2 q^{\frac{1}{2}} + 6 q + 12 q^{\frac{3}{2}} + 23 q^2
            + 42 q^{\frac{5}{2}} + 74 q^{3}
            + 124 q^{\frac{7}{2}} + 203 q^4 + 324 q^{\frac{9}{2}} \nonumber \\
&  \qquad \qquad  \qquad     + 502 q^5 + 768 q^{\frac{11}{2}}  + 1141 q^6 +1567 q^{\frac{13}{2}} + 1866 q^7
            + \cdots \Bigr) \\[4pt]
&\chi_{\rm 10A}(\tau) = q^{-\frac{1}{4}} \Bigl( 1 + 4 q^{\frac{1}{2}} + 7 q + 12 q^{\frac{3}{2}} + 24 q^2
            + 40 q^{\frac{5}{2}} + 63 q^{3}
            + 100 q^{\frac{7}{2}} + 153 q^4 +232 q^{\frac{9}{2}} \nonumber \\
&  \qquad \qquad  \qquad     + 342q^5 + 492 q^{\frac{11}{2}}  + 692 q^6 +900 q^{\frac{13}{2}} + 1018 q^7
            + \cdots \Bigr) \\[4pt]
&\chi_{\rm 11A}(\tau) = q^{-\frac{1}{4}} \Bigl( 1 + 2 q^{\frac{1}{2}} + 4 q + 4 q^{\frac{3}{2}} + 8 q^2
            + 14 q^{\frac{5}{2}} + 17 q^{3}
            + 24 q^{\frac{7}{2}} + 37 q^4 +52 q^{\frac{9}{2}} \nonumber \\
&  \qquad \qquad  \qquad     + 68q^5 + 88 q^{\frac{11}{2}}  + 116 q^6 +141 q^{\frac{13}{2}} + 147 q^7
            + \cdots \Bigr) \\[4pt]
&\chi_{\rm 12A}(\tau) = q^{-\frac{1}{4}} \Bigl( 1 + 4 q^{\frac{1}{2}} + 7 q + 12 q^{\frac{3}{2}} + 26 q^2
            + 48 q^{\frac{5}{2}} + 78 q^{3}
                        + 128 q^{\frac{7}{2}} + 211 q^4 +336 q^{\frac{9}{2}} \nonumber \\
&  \qquad \qquad  \qquad     + 516q^5 + 780 q^{\frac{11}{2}}  + 1157 q^6 +1592 q^{\frac{13}{2}} + 1886 q^7
            + \cdots \Bigr) \\[4pt]
&\chi_{\rm 14A}(\tau) = \chi_{\rm 14B}(\tau) = q^{-\frac{1}{4}} \Bigl( 1 + 3 q^{\frac{1}{2}} + \tfrac{11}{2} q
            + 8 q^{\frac{3}{2}} + \tfrac{33}{2} q^{2} +  29 q^{\frac{5}{2}} + \tfrac{87}{2} q^3
            + 68 q^{\frac{7}{2}} + 108 q^4 \nonumber \\
&  \qquad \qquad  \qquad   \qquad
+ 166 q^{\frac{9}{2}} + \tfrac{487}{2}q^5 + 352 q^{\frac{11}{2}}
+ \tfrac{1009}{2} q^6 + \tfrac{1335}{2}q^{\frac{13}{2}} + 761 q^7 + \cdots \Bigr) \\[4pt]
&\chi_{\rm 15A}(\tau) = \chi_{\rm 15B}(\tau) = q^{-\frac{1}{4}} \Bigl( 1 + 3 q^{\frac{1}{2}} + 6 q
            + 10 q^{\frac{3}{2}} + \tfrac{41}{2} q^{2} +  37 q^{\frac{5}{2}} + \tfrac{119}{2} q^3
            + 96 q^{\frac{7}{2}} + \tfrac{311}{2} q^4 \nonumber \\
&  \qquad \qquad  \qquad   \qquad
+ 244 q^{\frac{9}{2}} + 367 q^5 + 544 q^{\frac{11}{2}}
+ 795 q^6 + \tfrac{2149}{2}q^{\frac{13}{2}} + 1250 q^7 + \cdots \Bigr) \\[4pt]
&\chi_{\rm 21A}(\tau) = \chi_{\rm 21B}(\tau) = q^{-\frac{1}{4}} \Bigl( 1 + 4 q^{\frac{1}{2}} + \tfrac{15}{2} q
            + 14 q^{\frac{3}{2}} + \tfrac{57}{2} q^{2} +  50 q^{\frac{5}{2}} + \tfrac{165}{2} q^3
            + 134 q^{\frac{7}{2}} + 215 q^4 \nonumber \\
&  \qquad \qquad  \qquad   \quad
+ 340 q^{\frac{9}{2}} + \tfrac{1033}{2} q^5 + 772 q^{\frac{11}{2}}
+ \tfrac{2271}{2} q^6 + 1545 q^{\frac{13}{2}} + 1813 q^7 + \cdots \Bigr) \\[4pt]
&\chi_{\rm 23A}(\tau) = \chi_{\rm 23B}(\tau) = q^{-\frac{1}{4}} \Bigl( 1 + 3 q^{\frac{1}{2}} + 7 q
            + 14 q^{\frac{3}{2}} + 26 q^{2} +  43 q^{\frac{5}{2}} + \tfrac{149}{2} q^3
            + 124 q^{\frac{7}{2}} + \tfrac{383}{2} q^4 \nonumber \\
&  \qquad \qquad  \qquad   \quad
+ 296 q^{\frac{9}{2}} + \tfrac{909}{2} q^5 + 682 q^{\frac{11}{2}}
+ 985 q^6 + \tfrac{2639}{2} q^{\frac{13}{2}} + 1548 q^7 + \cdots \Bigr) \ .
\end{align}

\newpage

\begin{table}[ht]\centerline{\scalebox{.90}{
\rotatebox{90}{$\begin{array}{rrrrrrrrrrrrrrrrrrrrrrrrrrrrrrrrrrrrrr}
 1 & 2 & 3 & 4 & 5 & 6 & 7 & 8 & 9 & 10 & 11 & 12 & 13 & 14 & 15 & 16 & 17 & 18 & 19 & 20 & 21 & 22 & 23 & 24 & 25 & 26\\\hline
 {\rm 1A}  &  {\rm 2A} &  {\rm 3A} & {\rm  5A} &  {\rm 4B} &  {\rm 7A}
 &  {\rm 7B} &  {\rm 8A} &  {\rm 6A} &  {\rm 11A} &  {\rm15A} &  {\rm 15B}
 &  {\rm 14A} &  {\rm 14B} &  {\rm 23A}  &  {\rm 23B} &  {\rm 12B} &  {\rm 6B}
 &  {\rm 4C} &  {\rm 3B} &  {\rm 2B} &  {\rm 10A} &  {\rm 21A} &  {\rm 21B} &  {\rm 4A} &  {\rm 12A}\\\hline
 &    &   &   &  {\rm 2A} &   &
 &  {\rm 4B} &  {\rm 3A} &   &  {\rm 3A} &  {\rm 3A} &  {\rm 2A} &  {\rm 2A} &
 &  &   {\rm 6B} &  {\rm 3B} &  {\rm 2B} &   &   &  {\rm 2B} &  {\rm 3B} &  {\rm 3B} &  {\rm 2A} &  {\rm 6A}\\
   &   &   &   &   &   &   &   &  {\rm 2A} &  &  &  &  &  &  &  &  {\rm 4C} &  {\rm 2B} &   &   &   &   &   &   &   &  {\rm 3A}\\\hline
 1 & 1 & 1 & 1 & 1 & 1 & 1 & 1 & 1 & 1 & 1 & 1 & 1 & 1 & 1 & 1 & 1 & 1 & 1 & 1 & 1 & 1 & 1 & 1 & 1 & 1 \\
 23 & 7 & 5 & 3 & 3 & 2 & 2 & 1 & 1 & 1 & 0 & 0 & 0 & 0 & 0 & 0 & -1 & -1 & -1 & -1 & -1 & -1 & -1 & -1 & -1 &
   -1 \\
 252 & 28 & 9 & 2 & 4 & 0 & 0 & 0 & 1 & -1 & -1 & -1 & 0 & 0 & -1 & -1 & 0 & 0 & 0 & 0 & 12 & 2 & 0 & 0 & 4 & 1
   \\
 253 & 13 & 10 & 3 & 1 & 1 & 1 & -1 & -2 & 0 & 0 & 0 & -1 & -1 & 0 & 0 & 1 & 1 & 1 & 1 & -11 & -1 & 1 & 1 & -3 &
   0 \\
 1771 & -21 & 16 & 1 & -5 & 0 & 0 & -1 & 0 & 0 & 1 & 1 & 0 & 0 & 0 & 0 & -1 & -1 & -1 & 7 & 11 & 1 & 0 & 0 & 3 &
   0 \\
 3520 & 64 & 10 & 0 & 0 & -1 & -1 & 0 & -2 & 0 & 0 & 0 & 1 & 1 & 1 & 1 & 0 & 0 & 0 & -8 & 0 & 0 & -1 & -1 & 0 &
   0 \\
 45 & -3 & 0 & 0 & 1 & e_7^+ & e_7^- & -1 & 0 & 1 & 0 & 0 & -e_7^+ & -e_7^- & -1 & -1 & 1 &
   -1 & 1 & 3 & 5 & 0 & e_7^- & e_7^+ & -3 & 0 \\
 45 & -3 & 0 & 0 & 1 & e_7^- & e_7^+ & -1 & 0 & 1 & 0 & 0 & -e_7^- & -e_7^+ & -1 & -1 & 1 &
   -1 & 1 & 3 & 5 & 0 & e_7^+ & e_7^- & -3 & 0 \\
 990 & -18 & 0 & 0 & 2 & e_7^+ & e_7^- & 0 & 0 & 0 & 0 & 0 & e_7^+ & e_7^- & 1 & 1 & 1 & -1
   & -2 & 3 & -10 & 0 & e_7^- & e_7^+ & 6 & 0 \\
 990 & -18 & 0 & 0 & 2 & e_7^- & e_7^+ & 0 & 0 & 0 & 0 & 0 & e_7^- & e_7^+ & 1 & 1 & 1 & -1
   & -2 & 3 & -10 & 0 & e_7^+ & e_7^- & 6 & 0 \\
 1035 & -21 & 0 & 0 & 3 & 2 e_7^+ & 2 e_7^- & -1 & 0 & 1 & 0 & 0 & 0 & 0 & 0 & 0 & -1 & 1 & -1 & -3 &
   -5 & 0 & -e_7^- & -e_7^+ & 3 & 0 \\
 1035 & -21 & 0 & 0 & 3 & 2 e_7^- & 2 e_7^+ & -1 & 0 & 1 & 0 & 0 & 0 & 0 & 0 & 0 & -1 & 1 & -1 & -3 &
   -5 & 0 & -e_7^+ & -e_7^- & 3 & 0 \\
 1035 & 27 & 0 & 0 & -1 & -1 & -1 & 1 & 0 & 1 & 0 & 0 & -1 & -1 & 0 & 0 & 0 & 2 & 3 & 6 & 35 & 0 & -1 & -1 & 3 &
   0 \\
 231 & 7 & -3 & 1 & -1 & 0 & 0 & -1 & 1 & 0 & e_{15}^+ & e_{15}^- & 0 & 0 & 1 & 1 & 0 & 0 & 3 & 0 & -9 & 1 &
   0 & 0 & -1 & -1 \\
 231 & 7 & -3 & 1 & -1 & 0 & 0 & -1 & 1 & 0 & e_{15}^- & e_{15}^+ & 0 & 0 & 1 & 1 & 0 & 0 & 3 & 0 & -9 & 1 &
   0 & 0 & -1 & -1 \\
 770 & -14 & 5 & 0 & -2 & 0 & 0 & 0 & 1 & 0 & 0 & 0 & 0 & 0 & e_{23}^+ & e_{23}^- & 1 & 1 & -2 & -7 & 10 & 0
   & 0 & 0 & 2 & -1 \\
 770 & -14 & 5 & 0 & -2 & 0 & 0 & 0 & 1 & 0 & 0 & 0 & 0 & 0 & e_{23}^- & e_{23}^+ & 1 & 1 & -2 & -7 & 10 & 0
   & 0 & 0 & 2 & -1 \\
 483 & 35 & 6 & -2 & 3 & 0 & 0 & -1 & 2 & -1 & 1 & 1 & 0 & 0 & 0 & 0 & 0 & 0 & 3 & 0 & 3 & -2 & 0 & 0 & 3 & 0 \\
 1265 & 49 & 5 & 0 & 1 & -2 & -2 & 1 & 1 & 0 & 0 & 0 & 0 & 0 & 0 & 0 & 0 & 0 & -3 & 8 & -15 & 0 & 1 & 1 & -7 &
   -1 \\
 2024 & 8 & -1 & -1 & 0 & 1 & 1 & 0 & -1 & 0 & -1 & -1 & 1 & 1 & 0 & 0 & 0 & 0 & 0 & 8 & 24 & -1 & 1 & 1 & 8 &
   -1 \\
 2277 & 21 & 0 & -3 & 1 & 2 & 2 & -1 & 0 & 0 & 0 & 0 & 0 & 0 & 0 & 0 & 0 & 2 & -3 & 6 & -19 & 1 & -1 & -1 & -3 &
   0 \\
 3312 & 48 & 0 & -3 & 0 & 1 & 1 & 0 & 0 & 1 & 0 & 0 & -1 & -1 & 0 & 0 & 0 & -2 & 0 & -6 & 16 & 1 & 1 & 1 & 0 & 0
   \\
 5313 & 49 & -15 & 3 & -3 & 0 & 0 & -1 & 1 & 0 & 0 & 0 & 0 & 0 & 0 & 0 & 0 & 0 & -3 & 0 & 9 & -1 & 0 & 0 & 1 & 1
   \\
 5796 & -28 & -9 & 1 & 4 & 0 & 0 & 0 & -1 & -1 & 1 & 1 & 0 & 0 & 0 & 0 & 0 & 0 & 0 & 0 & 36 & 1 & 0 & 0 & -4 &
   -1 \\
 5544 & -56 & 9 & -1 & 0 & 0 & 0 & 0 & 1 & 0 & -1 & -1 & 0 & 0 & 1 & 1 & 0 & 0 & 0 & 0 & 24 & -1 & 0 & 0 & -8 &
   1 \\
 10395 & -21 & 0 & 0 & -1 & 0 & 0 & 1 & 0 & 0 & 0 & 0 & 0 & 0 & -1 & -1 & 0 & 0 & 3 & 0 & -45 & 0 & 0 & 0 & 3 &
   0
\end{array}$}}}\caption{The character table of the Mathieu group. The rows correspond to the
characters, while the columns describe the different conjugacy classes, numbered from $1$ to $26$
as in \cite{EOT}. The second line gives the name of the conjugacy class that is used throughout
the paper, where the number is the order of any element in the class. Where appropriate the
classes of some powers are given in the third and fourth row, but only the non-trivial cases are stated.
Finally, $e_p^\pm=(-1\pm i\sqrt{p})/2$.}
\end{table}

\end{document}